# Magnetic field concentration with coaxial silicon nanocylinders in optical spectral range

**Kseniia V. Baryshnikova**[*,1], **Andrey B. Evlyukhin**[1,2], **and Alexander S. Shalin**[1,3,4]

[1] Laboratory "Nanooptomechanics", ITMO University, 49 Kronverksky Pr., 197101 St. Petersburg, Russia

[2] Laser Zentrum Hannover e.V., Hollerithallee 8, D-30419 Hannover, Germany

[3] Kotel'nikov Institute of Radio Engineering and Electronics of Russian Academy of Sciences (Ulyanovsk branch), Goncharova str. 48/2, 432071 Ulyanovsk, Russia

[4] Ulyanovsk State University, Lev Tolstoy str. 42, 432017 Ulyanovsk, Russia

**Abstract** Possibility of magnetic energy accumulation inside silicon nanoparticles at the conditions of resonant optical responses is investigated theoretically. The magnetic field distributions inside silicon nanocylinders with and without coaxial through holes are calculated using full-wave numerical approach. It is demonstrated that such systems can be used for control and manipulation of optical magnetic fields providing their enhancement up to 26 times at the condition of optical resonances. Obtained results can be used for realization of nanoantennas and nanolasers, in which magnetic optical transitions play significant roles.

**1 Introduction** Concentration of light to small volumes is a topic of extensive investigations, suggesting a broad range of possible applications, e.g. [1–4]. While performances of conventional optical elements cannot lead to substantial field concentrations below classical diffraction limit, various auxiliary nanostructures, operating with near field, could deliver deep subwavelength enhancement. In particular, negative permittivity nanoparticles, supporting localized plasmon resonances, are capable of concentrating electric fields at nanoscale volumes [5]. Electro-magnetic energy in this case is stored in electrical field during one-half of an oscillation period, while electron dis-placement in negative epsilon materials accumulates the energy during its second half. Plasmonic particles concentrate fields at their close vicinity and, as the result, are widely employed for enhancing fluorescence [6], improving sensing [7], achieving nano-scale lasing [8], nano-positioning with optical tweezing [9–11] and for many other applications. An approach for manipulating magnetic field component of light relies on adopting magnetic Mie resonances in high refractive index particles [12–15]. In this case, circular displacement currents yet in fully retarded regime, create effective magnetic dipolar and higher or-der resonances [16]. Magnetic resonances of this kind were studied in various configurations, aiming to tailor optical properties of micro- and nanostructures by exploring and adjusting a set of their geometrical parameters. While spherical geometries have only 2 degrees of freedom: their radius and material parameters (silicon and germanium are the most frequently used), cylinders provide higher degree of flexibility in optical response tuning. The phenomenon in high-index particles relies on retardation effects, and, as the results, fields are attracted to regions containing higher permittivity materials, similar to the waveguiding phenomena [17]. Field concentrations in those geometries are achieved inside high refractive index materials and, as the result, are inaccessible by external probes. However, electromagnetic fields could be accessible in small gaps in-side high refractive index particles [18]. Note, that the magnetic responses of particles with hollows had been considered previously [14, 19–22], but only with the view of scattering problem, while magnetic field concentration had not been investigated. In this Letter we investigate the magnetic field distribution inside silicon nanocylinders with and without coaxial through holes. The conditions for obtaining high-efficient MHSs are investigated in details. Possibility for magnetic field accumulation and enhancement up to 26 times in free space is demonstrated. This effect can be subsequently employed for designing particle-based magnetic field concentrators.

The Letter is organized as follows: optical properties of silicon cylinders will be studied first. Optimization over geometrical parameters of cylinders for achieving high magnetic field concentrations will be performed. Impact of multipolar interference in overlapping resonances under plane wave excitation will be briefly discussed. Silicon nanocylinders geometry will be then changed to the coaxial layout on the way for achieving high magnetic field concentration in the air void. Optical properties of these structures will be investigated and will be subsequently shown to deliver relatively high magnetic local field enhancement. All the numerical calculations will be done with help of Finite Element Method, realized in Comsol Multiphysics [available at: www.comsol.com].

**2 Magnetic hot-spots in silicon nanocylinders** We consider a silicon cylinder excitation by a plane wave of linear polarization and two different propagation directions: along (frontal excitation) and perpendicular (lateral excitation) to the cylinder's axis (see fig.1A). The case, when propagation is perpendicular to the cylinder's axis and electric field oscillates along it, is very similar to the case of frontal excitation concerning magnetic hot-spots (MHSs).

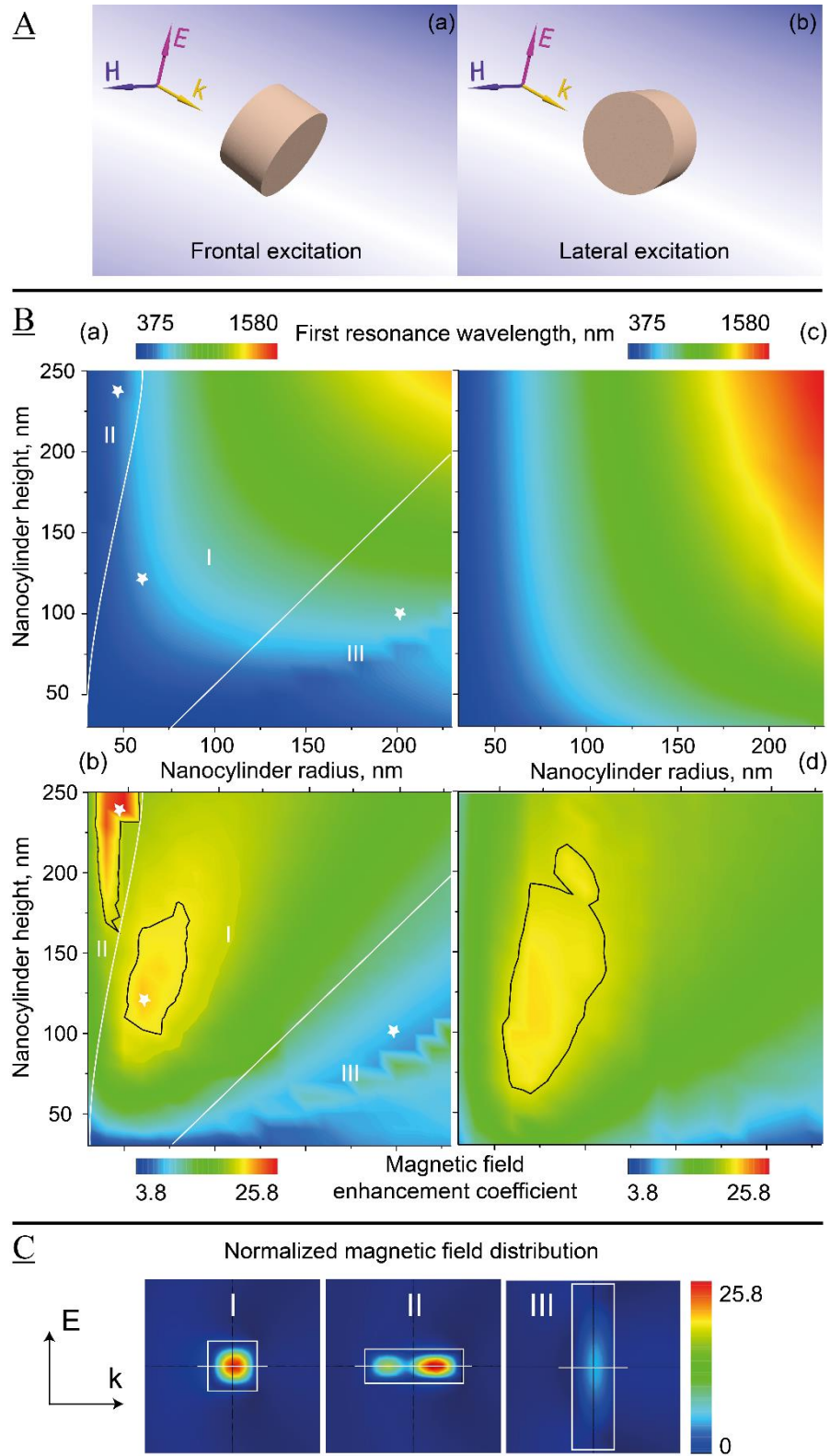

**Figure 1** A: Cylinders, irradiated by a plane wave, which (a) propagates along the cylinder axis (frontal excitation); (b) perpendicularly to the cylinder axis (lateral excitation). B: (a, c) Maps of the first resonance wavelength (longest resonance wavelength) corresponding to the maximal magnetic field on the cylinder's

axis as a function of cylinder size parameters: (a) for the frontal excitation; (c) for the lateral excitation. (b, d) Maps of magnetic field enhancement coefficient (normalized magnetic field maximum), at the first resonance (resonance, corresponding to the longest wave-length) as a function of cylinder size parameters: (b) for the frontal excitation, (d) for the lateral excitation. Three different areas, high-lighted by white lines (a, b), correspond to the different multipole orders responsible for the first resonance: I – first resonance is caused by magnetic dipole; II – first resonance is caused by the combination of magnetic dipole and magnetic quadrupole resonances; III – combination of several higher-order multipoles. White stars (a, b) correspond to the parameters, chosen for MHS illustration in fig.1C. Black contoured islands (b, d) highlight areas where the enhancement coefficient is greater than 20.   C: Normalized magnetic field distribution for the first resonance in nanocylinders in the case of the frontal excitation (side view). Parameters of the nanocylinders: I – radius $R = 60$ nm, height $H = 120$ nm; II – $R = 40$ nm, $H = 230$ nm; III – $R = 200$ nm, $H = 100$ nm. All cases are shown in fig. 1B (a, b) by white stars.

In our numerical simulations, in order to realize magnetic hot-spots inside the cylinders in the visible and near-infrared ranges, we varied the height and radius of cylindrical nanoparticles in the ranges 30-250 nm and 30-230 nm respectively [12]. We investigated spectral dependencies of the magnetic field maximum located on the cylinder's axis, and it was shown, that these dependencies can have several maxima depending on the cylinder geometry (see details further). So, let us firstly consider the first resonance, i.e. resonance, corresponding to the longest wave-length, independently on the order of corresponding multi-pole. This terminology seems to be more preferable for the describing magnetic field concentration because resonant MHS can correlate to the combination of the several multipoles depending on the geometrical parameters of nanoparticles. The high-order resonances will be studied further in this Letter.

The light wavelength of the first resonance and corresponding magnetic field maximum, normalized on the amplitude of the incident magnetic field, are shown in fig. 1B as a function of cylinder's height and radius. It can be seen that the resonant wavelength and the value of magnetic field on the first resonance are strongly dependent on the irradiation conditions.

Our multipole analysis (see [16] for details, the multi-pole decomposition formulas will not be shown here), shows that in the case of the frontal excitation (fig. 1A-a) the first optical resonance can correspond to the resonances of different multipole moments of the nanoparticles de-pending on the size parameters).  In fig. 1B-a, b the area I corresponds to the excitation of magnetic dipole as a first resonance, the area II corresponds to combination of magnetic dipole and magnetic quadrupole as a first resonance, and the area III corresponds to a combination of several electric and magnetic multipoles.

In the case when the first resonance and MHSs are caused by the magnetic dipole resonance (area I) the maximum of the magnetic field inside nanoparticles can exceed twenty times the magnetic field amplitude of an incident light (see black contour in fig. 1B-b). It is necessary to mention, that in the case of lateral irradiation (fig. 1A-b) of nanoparticles the first resonance always corresponds to magnetic dipole resonance independently on the particle's geometry (fig. 1B-c, d) in the considered range of parameters. Distribution of magnetic field enhancement coefficient in fig. 1B-d is similar to the distribution shown in the area I in fig. 1B-b.

For nanoparticles with size parameters from the area II, the value of enhancement coefficient can be even larger, up to 26 times (see fig. 1B-b). In the area III the magnetic hot-spots are relatively weak. Distributions' inhomogeneity in this area (see fig. 1B-a, b) is caused by the strong sensitivity of multipoles to the cylinder's aspect ratio [12]. Characteristic distributions of magnetic field inside nanoparticles from the areas I, II, III are shown in fig. 1C.

**3 Magnetic hot-spots in coaxial silicon nanocylinders with void core** Above we considered the magnetic hot-spots inside the nanoparticles volume. Meanwhile for practical applications it is important to have direct access to the MHSs. Several approaches have been suggested for these purposes previously. For example, MHS can be obtained in a small gap between high-index nanoparticles [23, 24] or in the through cavity [14, 22].

**3.1 Magnetic hot spots of first resonance** In the case of dielectric (silicon) nanocylinders, when strong magnetic field concentrates inside nanoparticle (fig.1C), they could be accessed by using a coaxial through hole (or cavity), see fig. 2a (frontal irradiation), 2b (lateral irradiation) [14, 22]. Such holes can be produced by focused ion-beam milling or electron beam lithography and reactive ion etching [14, 22]. In fig. 2 the results of numerical simulations for cylinders with radius $R = 60$ nm and height $H = 120$ nm with through holes are presented for the both types of irradiation. Note, that these geometrical parameters correspond to the maximal magnetic field enhancement up to 22 times in fig. 1B-d, which is caused by magnetic dipole resonance. In the case of magnetic dipole resonance similar dependencies (as in fig. 2) take place for nanocylinders with other parameters considered above.

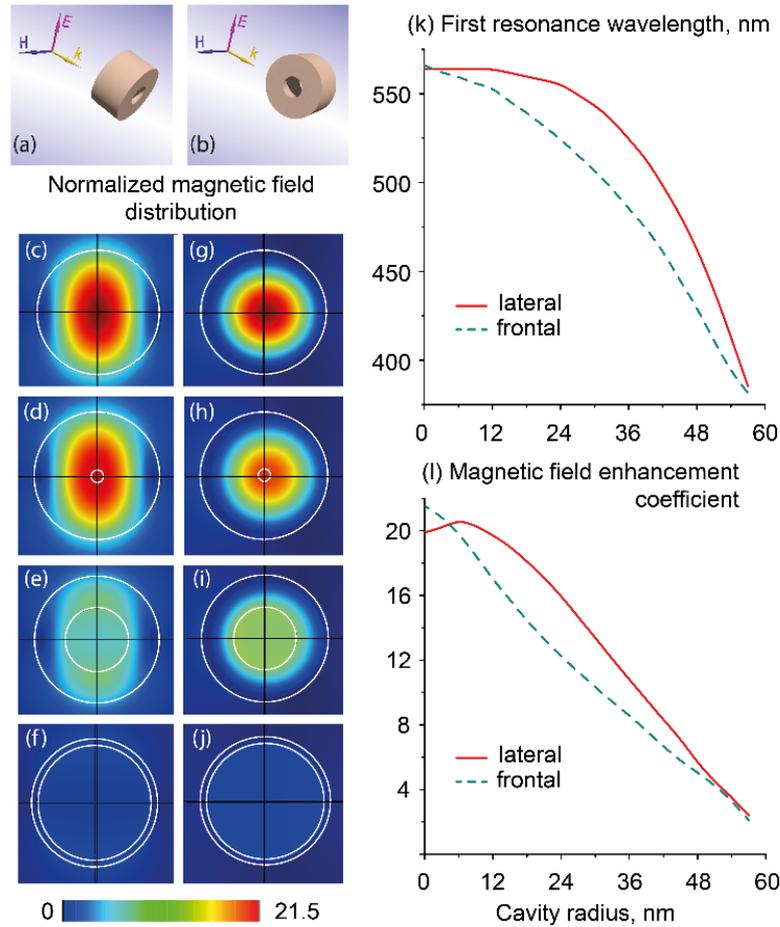

**Figure 2** (a, b) Frontal (a) and lateral (b) excitation of cylinder with a coaxial through hole. (c-j) Distributions of normalized magnetic field for the first resonance in the cylinders with radius R = 60 nm and height H = 120 nm (these parameters correspond to the maximal magnetic field enhancement in fig. 1B-d) and cylinders with coaxial hole of different radii. (c-f) - frontal and (g-j) - lateral excitation. (d) cavity radius is 6 nm, resonant wavelength is 566 nm; (e) cavity radius is 30 nm, resonant wavelength is 507 nm; (f) cavity radius is 54 nm, resonant wavelength is 494 nm; (h) cavity radius is 6 nm, resonant wavelength is 564 nm; (i) cavity radius is 30 nm, resonant wavelength is 543 nm; (j) cavity radius is 54 nm, resonant wavelength is 440 nm. (k) The wavelengths corresponding to the first resonance as a function of hole radius. (l) Magnetic field enhancement coefficient in the hole as a function of its radius.

Distribution of magnetic field in particles with small cavity is nearly the same as for nanoparticle without cavity (fig. 2c-d, g-h). Magnitude of magnetic field is quite homogeneous on the frontal cross-section and has a Lorentz-like maximum near the middle of cylinder's axis. The magnetic field in the cavity becomes weaker with in-creasing the cavity size (see fig. 2d-f, h-j). However, even for silicon shell of 6 nm thickness the magnetic field magnitude in the cavity is enhanced up to 2 times for the both frontal and lateral irradiation conditions. When the radius of cavity is larger than 6 nm, the lateral excitation of nanocylinders is more prospective for realization strong magnetic fields (fig. 2l). It origins from the fact that the circular character of displacement currents is not changed for the lateral irradiation. In the case of the frontal excitation the circular currents are discontinuous that leads to the rapid destruction of the magnetic hot-spots with increasing of the cavity radius. The wavelength corresponding to the MHS excitation decreases monotonically when size of cavity goes up in both cases (fig. 2k). Importantly, that the dependencies presented in fig. 2k, l allow us to con-struct the system with required levels of the magnetic field enhancement in a through hole for a certain spectral range.

### 3.2 Magnetic hot-spots of higher-order resonances

Now, let us consider the magnetic field enhancement on the high-order resonances with wavelengths lower than the wavelength of the first resonance [26]. The possible spectral dependencies of magnetic field enhancement coefficient are presented in fig. 3 for frontal (a) and lateral (b) excitation. Insets show the distributions of the normalized magnetic field, corresponding to the resonances.

The volume, where magnetic field is concentrated, decreases for high-order resonances comparing with the first resonance and MHSs can split into two or more hot-spots along the cylinder's axis (e.g. "x2" resonances in fig. 3). For the lateral excitation, the magnetic field enhancement is larger for the first resonance, whereas, for the frontal excitation the second resonance provides stronger MHSs with magnetic field enhancement coefficient up to 25 times. Additional calculations show similar dependencies for the cylinders from the considered range of parameters, how-ever, for small cylinders high-order resonances can move into shorter wavelengths region or even disappear because of the strong changes of silicon refractive index in ultraviolet spectral range.

Higher-order resonances correspond to more inhomogeneous distributions of displacement currents inside nanocylinders. Therefore, these resonances are very sensitive to the cylinder's defects like a cavity or coaxial hole. For example, some resonances, marked by the "*" index on the fig. 3, disappears for cavity radius > 20 nm due to the redistribution of energy to the MHSs inside the nanoparticle volume (see insets in Fig.3).

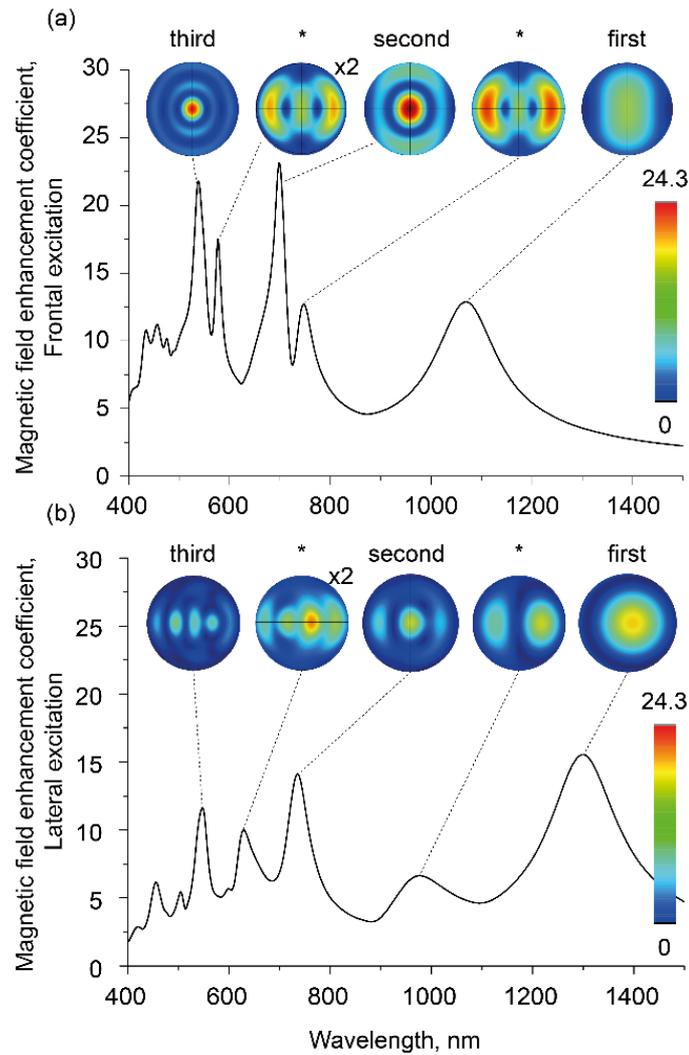

**Figure 3** Spectra of magnetic field enhancement coefficient for the cylinder with $R = 190$ nm, $H = 180$ nm without hollow for (a) frontal excitation, and (b) lateral excitation. Insets show distributions of normalized magnetic field at the resonance wavelengths at the plane crossing MHS center, where magnetic field is maximal. Index "x2" corresponds to the two maxima on the cylinder's axis. Index "*" corresponds to the resonances, which are destroyed in the cylinders with cavity of radius > 20 nm. All the insets correspond to the cylinders without cavity.

Dependencies of the resonant wavelengths and the magnetic field enhancement coefficients on the resonances, marked on fig. 3 as "first", "second", and "third" are shown in fig. 4a, b (for the frontal excitation) and c, d (for the lateral excitation), respectively. The magnetic field enhancement coefficient and resonant wavelength are changed with the variation of cavity size, and these dependences are much more pronounced for the high-order resonances (fig. 4).

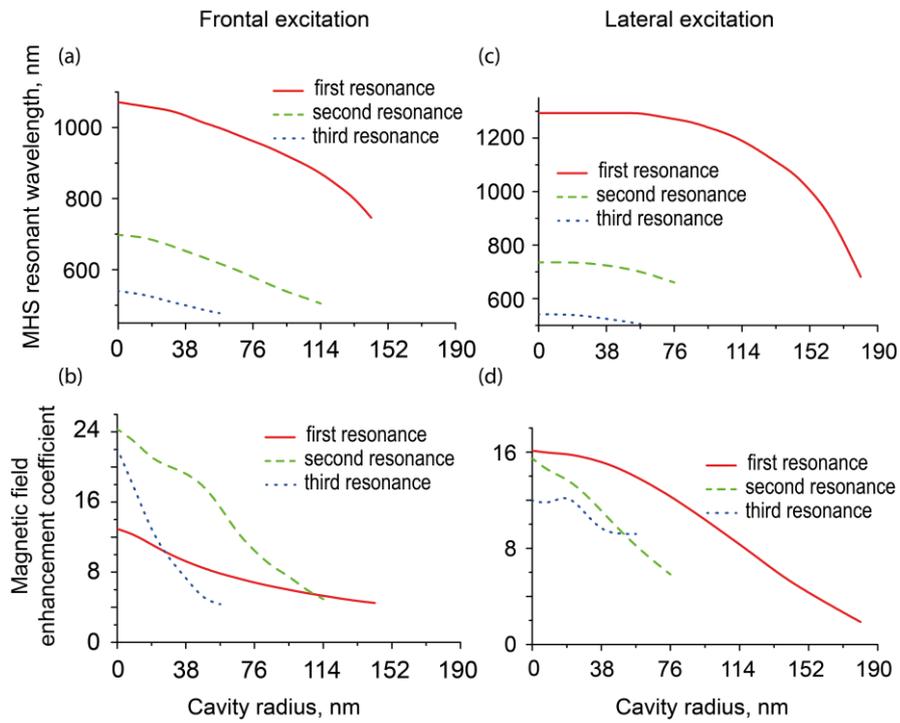

**Figure 4** (a, b) Dependencies of the resonant wavelengths and magnetic field enhancement coefficients on the cavity radius for the "first", "second" and "third" resonances (see fig. 3) in the case of frontal excitation. (c, d) – the same, but for lateral excitation. All the results are obtained for cylinder with R = 190 nm, H = 180 nm.

The "third" resonance's MHS disappears totally for the cavity of 50 nm radius independently on the irradiation conditions. The "second" resonance's MHS disappears for cavity radius > 114 nm in the case of the frontal, and > 76 nm in the case of the lateral irradiation. These dependencies could be useful for construction of the magnetic field's management system.

Finally, in practical applications nanoparticles are usually placed on some substrates. When refractive index and absorption coefficient of a substrate are small (as a glass), its influence on the MHS properties is rather insignificant [27]. However, in the case of high-index or high-absorbing substrates the interaction between particle and substrate plays an important role [14, 28, 29].

As a conclusion, we have investigated the magnetic field distribution inside silicon nanocylinders with and without coaxial through holes using full-wave numerical simulations. Two cases of the irradiation conditions have been included: 1) along and 2) perpendicular to the cylinder axis. It has been found that in the both cases the high-efficient MHSs corresponding to different cylinder multi-pole moments could be excited. The conditions for obtaining MHSs have been investigated in details. The system is appeared to be quite tunable and allowed to control and manipulate the values of magnetic field enhancement and resonant wavelengths in broad spectral ranges. These results could be useful for different scientific areas such as nanophotonics and nanooptomechanics, where MHSs can be used for management, trapping and detection of magnetic nanoparticles and molecules with magnetic transitions.

**Acknowledgements** We are very grateful to C. R. Simovski and P. B. Ginzburg for fruitful discussions. This work has been supported by the Russian Fund for Basic Research within the project 16-52-00112. The calculations of magnetic field distributions and multipole moments has been supported by the Russian Science Foundation Grant No. 16-12-10287. A.S. acknowledges the support of the President of Russian Federation in the frame of Scholarship SP-4248.2016.1 and the support of Ministry of Edu-cation and Science of the Russian Federation (GOSZADANIE 2014/190). K.B. acknowledges the support of FASIE.